\documentclass[pre,aps,superscriptaddress,showpacs]{revtex4}

\usepackage{amsmath}
\usepackage{graphicx}

\newcommand{\beq}{\begin{equation}}
\newcommand{\eeq}{\end{equation}}
\newcommand{\be}{\begin{equation}}
\newcommand{\ee}{\end{equation}}
\newcommand{\bea}{\begin{eqnarray}}
\newcommand{\eea}{\end{eqnarray}}
\newcommand{\barr}{\begin{array}}
\newcommand{\earr}{\end{array}}

\begin{document}

\title{Dynamical properties of vibrofluidized granular mixtures}

\author{D.Paolotti}
\affiliation{
Dipartimento di Fisica \\
Universit\'a di Camerino, 62032 Camerino, Italy}
\affiliation{
Istituto Nazionale di Fisica della Materia, Unit\`a di Camerino, Camerino, Italy}
\author{C. Cattuto}
\affiliation{Frontier Research System, The Institute of Physical and Chemical
Research (RIKEN), Wako-shi, Saitama 351-0198, Japan and
Istituto Nazionale di Fisica della Materia, Unit\`a di Perugia}
\author{U. \surname{Marini Bettolo Marconi}}
\affiliation{
Dipartimento di Fisica \\
Universit\'a di Camerino, 62032 Camerino, Italy}
\affiliation{
Istituto Nazionale di Fisica della Materia, Unit\`a di Camerino, Camerino, Italy}
\author{A. \surname{Puglisi}}
\affiliation{
Dipartimento di Fisica \\
Universit\'a ``La Sapienza'',P.le A. Moro 2, 00198 Roma, Italy}
\affiliation{
Istituto Nazionale di Fisica della Materia, Unit\`a di Roma, Roma, Italy}

\date{\today}

\pacs{45.70.-n,05.40.-a,81.05.Rm}

\begin{abstract}
Motivated by recent experiments we have carried out an Event Driven
computer simulation of a diluted binary mixture of granular particles
vertically vibrated in the presence of gravity. The simulations not only
confirm that the kinetic energies of the two species are not equally
distributed, as predicted by various theoretical models, but
also seem to reproduce rather well the density and temperature
profiles measured experimentally. Rotational degrees of freedom 
do not seem to play any important qualitative role.
Instead, simulation shows the onset of a clustering instability
along the horizontal direction.
At the interior of the cluster we observe a secondary instability with
respect to the perfect mixing situation,
so that segregation of species is observed within the cluster.

\end{abstract}

\maketitle


\section{Introduction}

The present keen interest in the dynamical properties of granular
materials is motivated both by the challenge of understanding the
complex processes involved and by the important practical
applications in engineering, industry and technology
\cite{general}.  These materials are peculiar in many respects and
display several intriguing phenomena such as clustering
\cite{Zanetti}, shear instability
\cite{Brey,Baldassa} and lack of energy equipartition, 
which make their behavior 
different from ordinary molecular fluids.
The dissipation of kinetic energy during the inelastic collisions
makes them special.
The main motivation of the present paper stems from two recent experiments
\cite{Feitosa},\cite{Parker} which demonstrated 
that when a mixture constituted by two different species of 
grains is vibrated, each component attains its own ``granular temperature'',
i.e. the average kinetic energy per particle does not take on the
universal value $f K T$, where $f$ is the number of degrees of freedom
and $K$ a constant, as it occurs in molecular gases. On
the contrary, one observes that the ratio $T_1/T_2$ varies with
concentration, inelasticity parameters, particle sizes, masses and
driving mechanism.  Even in the absence of energy injection,
the inelastic gas cools, but one observes that the temperature 
ratio asymptotically remains constant.  On the other hand, 
it is understood that while the
only relevant hydrodynamic field is the global temperature $T=x T_1
+(1-x) T_2$, transport properties depend on that ratio \cite{dufty2}.

These are of course manifestations of the non-equilibrium nature of
Granular systems, which can only be maintained stationary by a
continuous energy feeding to compensate the energy losses due to the
inelastic collisions and to friction.

A theoretical understanding of such a behavior of granular mixtures
has been achieved in the case of homogeneous driving mechanisms by
means of a combination of models and approximations including the
pseudo-Maxwell inelastic gas and the Inelastic Hard Sphere model
treated by means of the Boltzmann-Enskog equation.  Both models have
been studied analytically and numerically in the free cooling
\cite{dufty,ioepuglio} and in the driven case
\cite{ioepuglio2,Pagnani,Barrat,barrat2}. Apart from the studies
of ref \cite{Pagnani,barrat2} none of these investigations considered
the role played by the gravitational field, by the strongly
inhomogeneous boundary conditions employed in the experiments of
refs.~\cite{Feitosa,Parker}, by the roughness of the grains and by
their rotational degrees of freedom. Moreover, there is still an open
debate about the ``best'' energy feeding mechanism. Whereas
theoreticians seem to favour a uniform thermal gaussian bath, because
it lends itself to a great deal of analytical work, a numerical
computer experiment can test directly driving mechanisms which are
closer to those employed in a laboratory.  The structure of the paper
is the following: in section 2 we illustrate briefly the model,
leaving the technical details to the appendix; in section 3 we discuss
the results for the geometry of ref.~\cite{Feitosa}. In sec. 4 we
consider a different aspect ratio, namely a longer box, where gravity
plays a more relevant role. Finally in sec. 5 we present our conclusions.


\section{Model system}

We decided to remain as close as possible to a commonly employed
experimental set-up, by constraining the grains to move on a vertical
rectangular domain of dimensions $L_x \times L_y$.  The gravitational
force acts along the negative $y$ direction.  The grains are assumed
to be spherical and free to rotate about an axis normal to the $xy$
plane.  They receive energy by colliding with the horizontal walls,
harmonically vibrating at frequency $\nu$.  The side walls instead are
immobile and were chosen either smooth or rough according to the
numerical experiment. When  side walls are considered to be rough,
they 
are assumed to have the same friction coefficient
$\mu$ as the particles.

The collisional model adopted in the present paper corresponds to the
one proposed by Walton \cite{Walton}. It conserves both the linear and
the angular momentum of a colliding pair, but allows energy to be
dissipated by means of a normal restitution coefficient and a
friction coefficient $\mu$.  The collision rule (given in detail in
the appendix) takes into account a reduction of normal relative
velocity of the two particles ($V_n$), a reduction of total tangential
relative velocity ($V_r$) and an exchange of energy between those two
degrees of freedom. The reduction of normal relative velocity is
modeled by means of a non constant restitution coefficients
$\alpha_{ij} \in [0,1]$, whose dependence by the relative velocity is
of the form:

\begin{displaymath}
\alpha_{ij} (V_n) = \left\{ 
\begin{array}{lll}
1 - (1 - r_{ij}) \left( \frac{|V_n|}{v_0} \right)^\frac{3}{4} & \mbox{ for } & V_n < v_0 \\
                                        r_{ij} & \mbox{ for } & V_n > v_0 
\end{array}
\right.
\end{displaymath}
where $i$ and $j$ are the numbers indicating the species of the
colliding particles, $r_{ij}$ are constants related to the three types
of colliding pairs, $v_0 \approx \sqrt{gd}$, where $d$ is the average
diameter of the particles and $g$ the gravitational acceleration
\cite{Bizon}.

Simulated collisions are of two types: with sliding or sticking point
of contact. When the following condition is satisfied (high relative
tangential velocity), the collision happens in a sliding fashion,
otherwise it is sticking:

\begin{equation}
\frac{|V_r|}{V_n} \geq \frac{l+1}{l}\mu(1 + \alpha_{ij})
\end{equation}

where $l$ is the dimensionless moment of inertia (equal to $2/5$ for
spheres), while $\mu$ is a static friction coefficient characterizing
the surface roughness of particles, assumed equal to the dynamical
friction coefficient.  The full dynamics consists of interparticle
collisions, and wall-particle collisions.  The trajectories between
collision events are parabolic arcs due to the presence of the
gravitational field.

An efficient Event Driven (ED) simulation 
code was employed to evolve the system~\cite{Cattuto}.

The two species were chosen to be spheres of equal diameters
$d=0.16cm$, and unequal masses $m_1=1.58 \cdot 10^{-2}g$ and $m_2=5.21
\cdot 10^{-3}g$, respectively.  The driving frequency was set to $50$
Hz, the vibration amplitude $A=3.5$ diameters so that the
corresponding dimensionless acceleration $\Gamma=A\omega^2/g =56 $.

All the averages quantities reported in the following have been
obtained by employing equally spaced data points separated by time
intervals $\Delta t=10^{-1}$ s in order to assure statistical
independence of measures. We performed a number $n=1.5\cdot10^5$ of
vibration cycles.

Barrat and Trizac~\cite{barrat2} have recently considered one of the
systems studied in the present paper. However, our treatment presents
some differences:

\begin{itemize}
\item
The bottom and top walls of~\cite{barrat2} move in a sawtooth manner
with a negligible excursion so that their positions are considered
fixed, while our walls move sinusoidally with a non-negligible
amplitude.
\item
The walls are not smooth in our treatment, but have a 
friction coefficient $\mu>0$.
\item
The collisional models of our treatment and that of Barrat-Trizac are
different (they do not distinguish between sticking and sliding
collisions).
\item
We take into account gravity.
\end{itemize}
\begin{figure}
\includegraphics[clip=true,height=10.0cm, keepaspectratio,angle=90]
{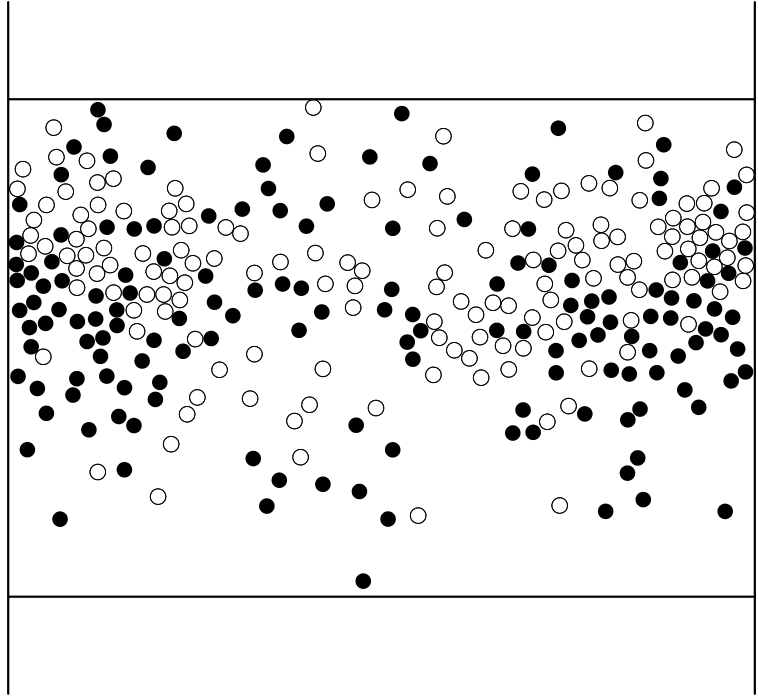}
\caption
{Typical snapshot of the system described in III.
Open circles indicate particles of species 1, black circles particles
of species 2.}
\label{snapshot1}
\end{figure}

\begin{figure}
\includegraphics[clip=true,width=10.0cm, keepaspectratio,angle=0]{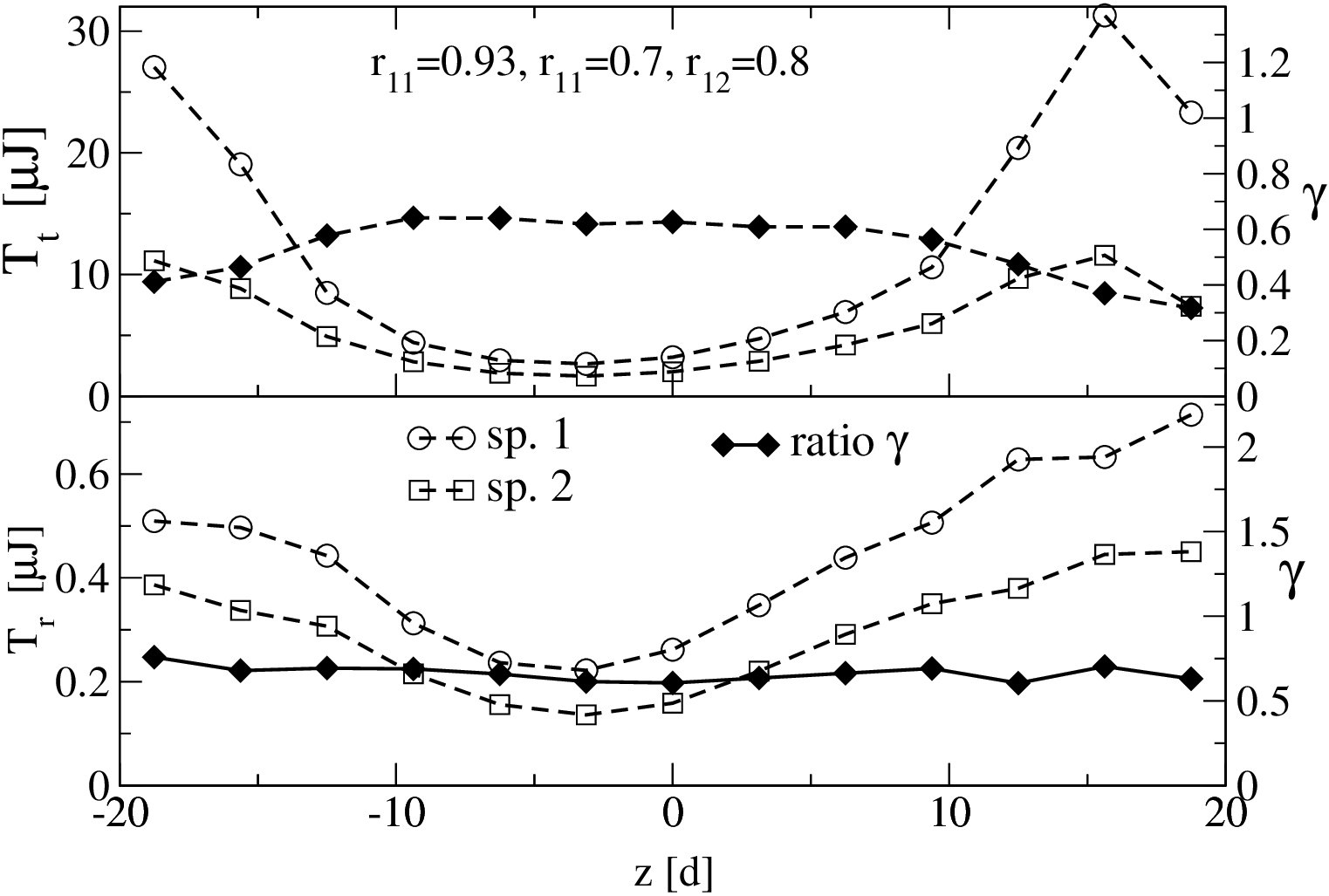}
\caption
{Top panel: Translational temperature profiles 
for species 1 (circles) and species 2 (squares) 
for a mass ratio $m_2/m_1=3.03$, expressed in $\mu J$ (left scale),
and temperature ratio (diamonds) $T_2/T_1$ (right scale).
The vertical position is measured in particle diameters ($d$)
relative to the geometric center of the cell.
Bottom panel: rotational temperature profiles 
for species 1 (circles) and species 2 (squares) 
and temperature ratio $T_2/T_1$ (diamonds).}
\label{tempA}
\end{figure}

\begin{figure}
\includegraphics[clip=true,width=10.0cm, keepaspectratio,angle=0]{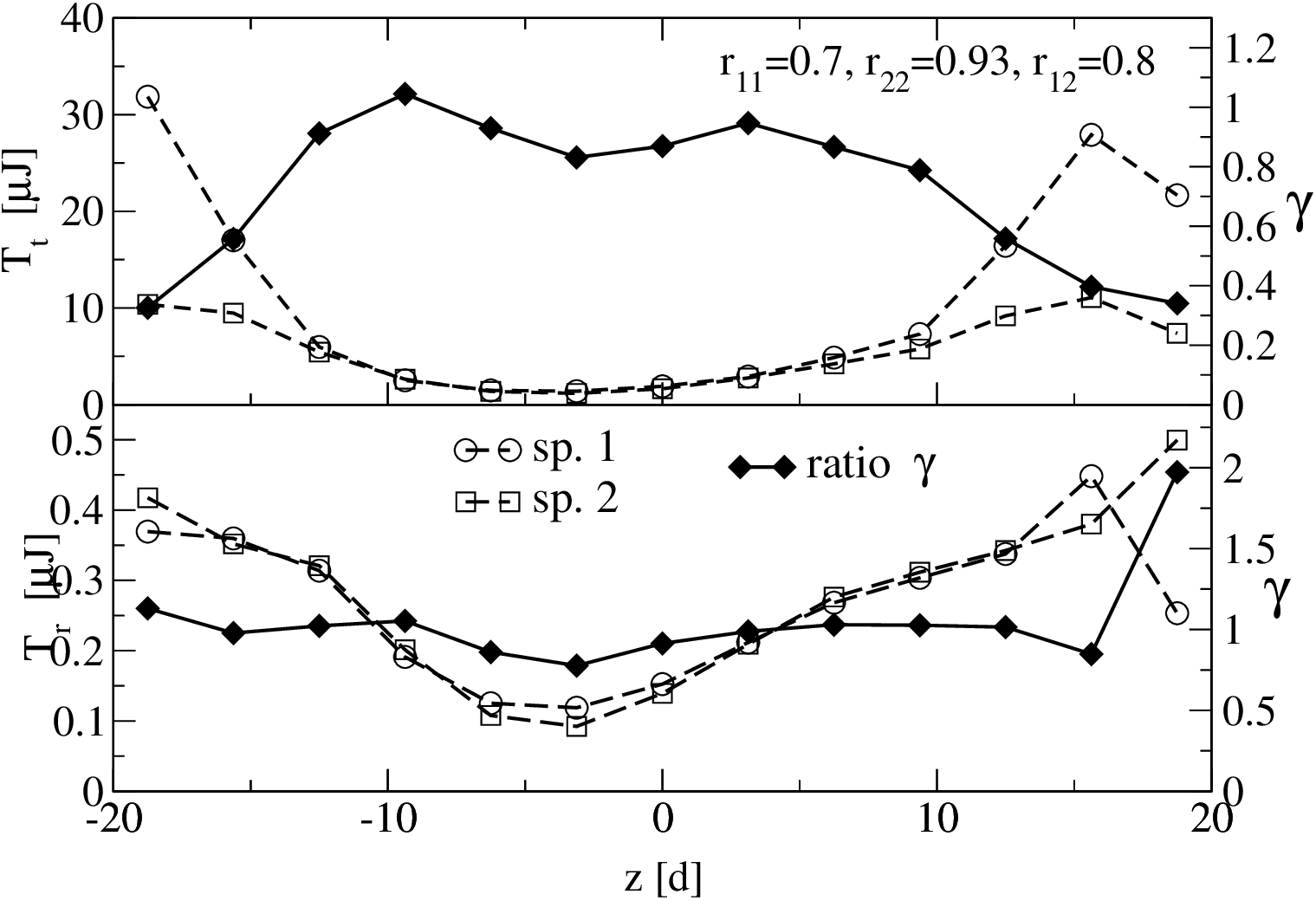}
\caption
{Same as in fig.1, but with different restitution coefficients.}
\label{tempB}
\end{figure}


\section{Results for a short system}

The box dimensions were $L_x=48\,d$  and $L_y=32\,d$.  Simulation runs
were carried out using $N_1=150$ grains of each species. The static
friction coefficient has been always chosen as $\mu=0.1$.  The
stationary state is determined by the balance between the energy input
provided by the vibrating walls at a frequency of $50$ Hz and the
dissipation due to inelastic collisions.  The typical collision
frequencies are of the order of $\nu_1 \sim 580$ Hz and $\nu_2 \sim
850$ Hz for the heavy and light balls, respectively. A typical 
microscopic configuration of the system is shown in fig.\ref{snapshot1}. 
\subsection{Temperature profiles}

\begin{figure}
\includegraphics[clip=true,width=10.0cm, keepaspectratio,angle=0]{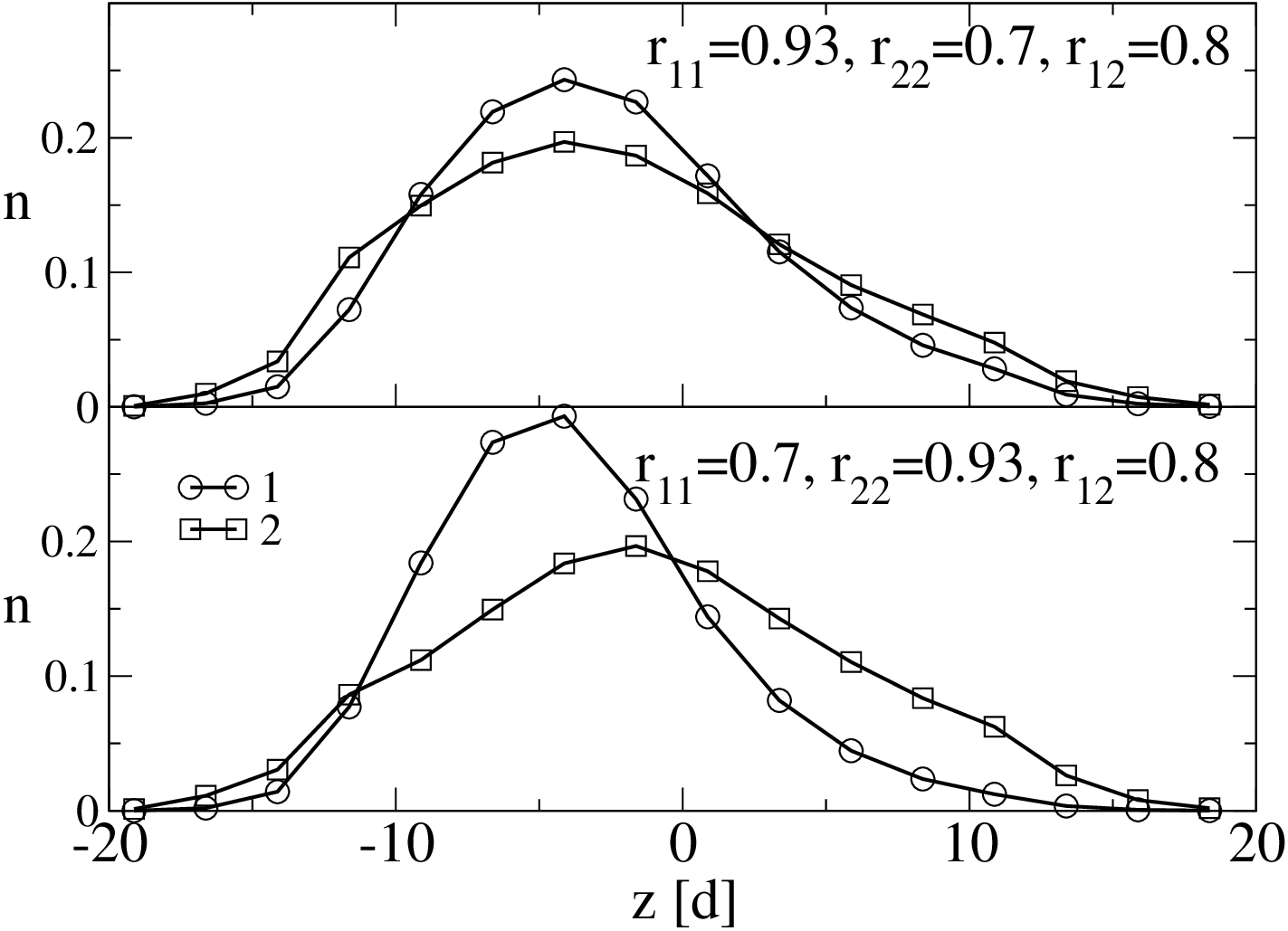}
\caption
{Area fraction profiles $n_1$ and $n_2$.  Control parameters are
the same as in fig. 1.}
\label{dens}
\end{figure}

The restitution coefficients were first set to $r_{11}=0.93$ for 1-1
collisions, $r_{22}=0.7$ for 2-2, and $r_{12}=0.8$ for 1-2
collisions. This means that the more massive particles are also the
more elastic ones. In figure~\ref{tempA} we show the partial
translational temperature profiles for the two species, and observe
that close to the vertical boundaries the two temperatures are
essentially determined by the energy injected by the vibrating
walls. Indeed, interparticle collisions are rare within this region,
and play no signifcant role because of the low local density (see
figure~\ref{dens}).  In addition, grains 1 and 2 impinging with the
same speed on the mobile wall bounce with the same velocity ($V\propto
A \omega$), hence the local value of the temperature ratio,
$\gamma=T_2/T_1$ near the vibrating walls, turns out to be
approximately $\gamma \sim m_1/m_2$, as shown in figure~\ref{tempA}.

On the other hand, the temperature drops as the distance from the
walls increases, while the ratio $\gamma$ grows up to a plateau value,
indicating that collisions tend to cool the mixture and render the two
partial temperatures closer. Figure~\ref{tempA} clearly displays the
breakdown of the kinetic energy equipartition already noticed in
previous experimental and theoretical studies.

We also measured the rotational temperature profiles, shown in
figure~\ref{tempA}. We observe for rotational
temperatures the same kind of equipartition breakdown that holds for
translational degrees of freedom.  Moreover, the absolute values of
rotational and translational temperatures are quite different, as
already reported by Luding \cite{Luding} for a one-component, vibrated
granulate.  On the other hand, the ratio of the two rotational
temperature profiles seems to be quite close to that of the
translational temperature profiles.

In figure~\ref{tempB}, we changed the restitution coefficients and set
$r_{11}=0.7$, $r_{22}=0.93$ for 2-2, and $r_{12}=0.81$. Therefore now
the more massive particles are the more inelastic. We observe that,
whereas the temperature profiles near the vibrating planes are nearly
unchanged, because collisions are rare, the value of the
temperature of the heavier species is lower and the temperature ratio
is closer to $1$. In this case the larger inelasticity of the heavier
particles competes with the mass asymmetry which instead tends to make
$T_2/T_1$ smaller.

\subsection{Area fraction profiles}

Due to the large value of the parameter $\Gamma$, the partial density
profiles tend to be rather symmetric, with a maximum near the center.
We also notice small differences in the density profiles, which reveal
that the heavier species has higher concentration close to the center
of the cell, while the lighter species is more spread.

Comparing the present results for the area fraction profiles with those
recently obtained by employing a Direct Simulation Monte Carlo technique
\cite{Pagnani}, we notice that the agreement is only qualitative,
whereas the temperature ratios are in significantly better agreement.

We verified that removing tangential friction (coupling the
translational and rotational degrees of freedom) does not change
significantly the above scenario. In fact, the translational
temperature and density profiles of the cases $\mu=0$ and $\mu=0.1$
show only small quantitative differences, more pronounced close to the
horizontal walls.

\begin{figure}
\includegraphics[clip=true,height=10.0cm, keepaspectratio,angle=-90]{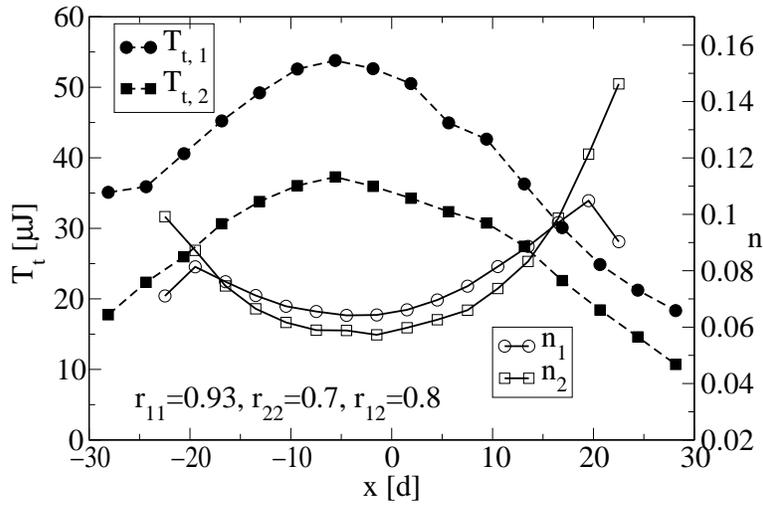}
\caption
{ Temperature profiles $T_1$ e $T_2$ (left scale) along the horizontal
direction and area fractions $n_1$ and $n_2$ (right scale). Control
parameters are the same as in fig. 1.}
\label{prof_xA}
\end{figure}  

\begin{figure}
\includegraphics[clip=true,height=10.0cm, keepaspectratio,angle=-90]{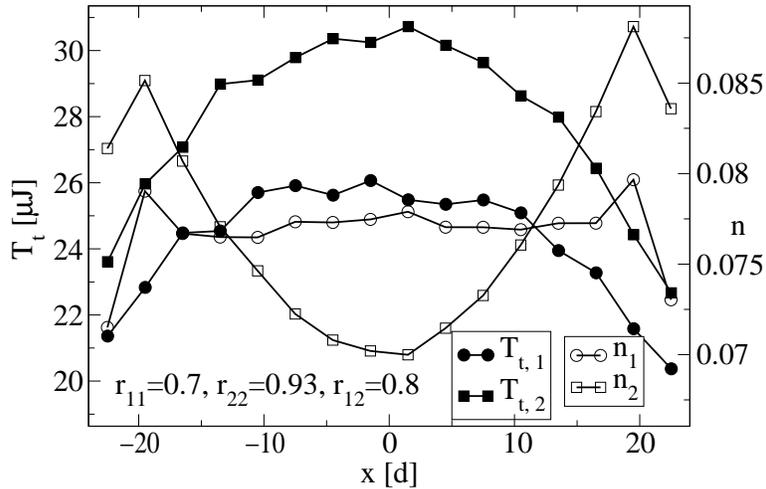}
\caption
{Same as in figure~\ref{prof_xA}, but with different restitution coefficients.}
\label{prof_xB}
\end{figure}  

\subsection{Transversal profiles}

We also studied the density and temperature profiles along the
horizontal direction.  To the best of out knowledge, no such measure
has been reported in experimental works. In figures~\ref{prof_xA}
and~\ref{prof_xB} we observe that temperature profiles vanish close to
side walls, while density profiles display their maxima in the same
region. In order to gain further insight, we analyzed a sequence of
snaphsots of the dynamics, and observed that the system bears a denser
cloud of grains in the vicinity of one of the side walls.
Such a configuration was maintained over an interval of time much longer
than the vibration period $2 \pi/\omega$. 

The cluster eventually ``evaporates'' to form again close to a randomly
selected side wall.  Over several periods of oscillation of the cell,
we noticed an effective horizontal simmetry breaking, i.e. the number
of particles in the right-hand and left-hand sides of the cell were
rather different.  The system is unstable with respect to horizontal
density fluctuations and clusterizes spontaneously, until the
vibrating bases wash out the cluster.

Moreover, we also observed some spontaneous tendency of the system to
segregate the species, a fact which becomes more apparent at high
densities.  Both phenomena have their origin in the inelasticity.  A
possible qualitative explanation of the observed dynamics is as
follows: particles with smaller restitution coefficient tend to group
together, since the more energy they dissipate through 2-2 collisions,
the denser the segregated domain becomes.  Particles with
higher restitution coefficient bounce for a longer time after a
collision, and dilate more quickly.

\subsection{Velocity Probability Distributions}

\begin{figure}
\includegraphics[clip=true,width=10.0cm, keepaspectratio,angle=0]{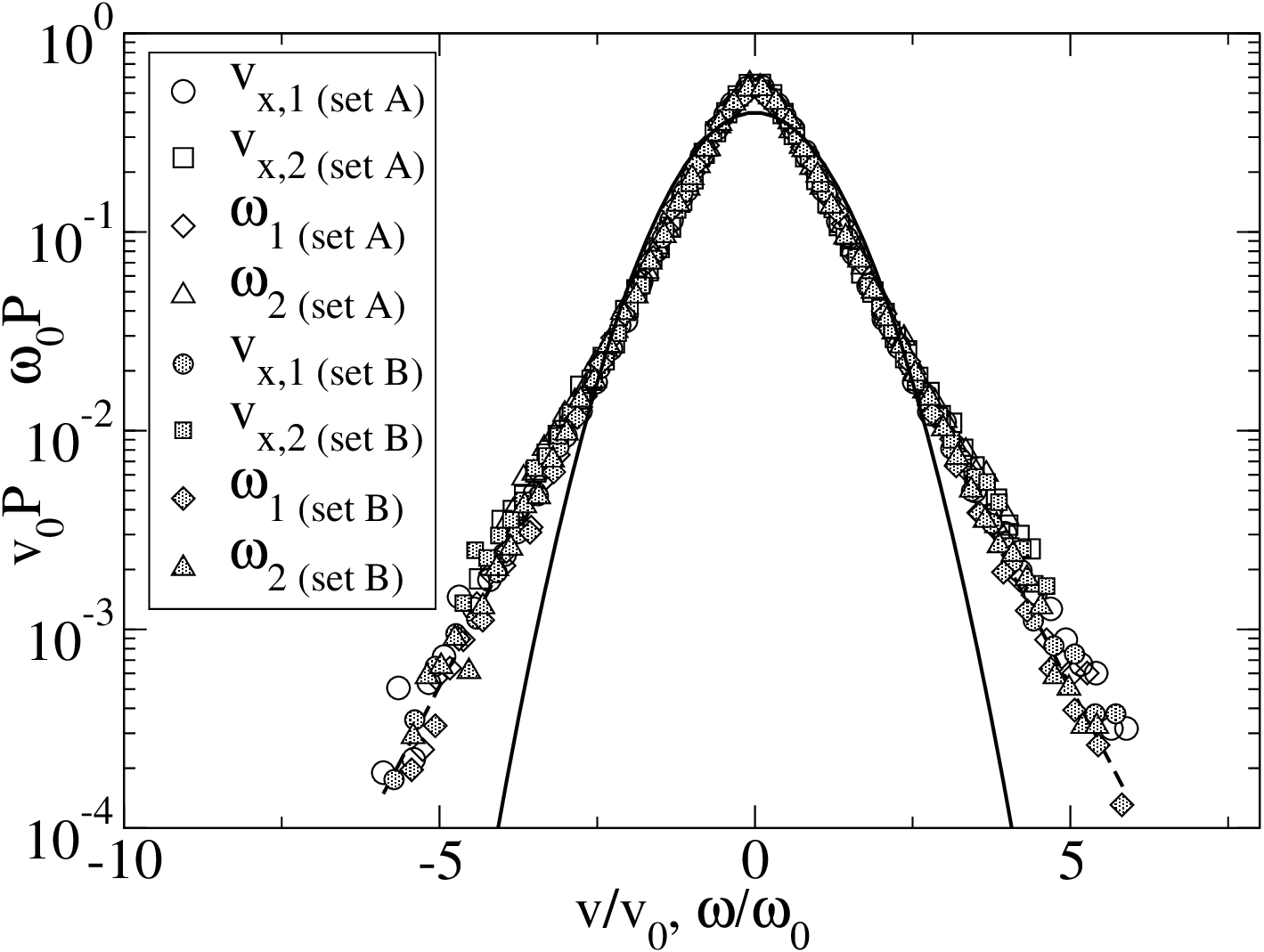}
\caption
{Rescaled velocity distribution functions $<c> P(c)=f(c/<c>)$ for
translational velocities and angular velocities of both species.  
Set A corresponds to parameters of fig. 2, set B to parameters of
fig. 3. 
The
independent variables ($v$, $\omega$) were rescaled by their mean
squared value ($v_0$, $\omega_0$). On rescaling, the distributions
collapse nicely onto each other. The dashed curve is the fitting law
discussed in the text, the continuous line is the Gaussian plotted as
a reference for the eye.}
\label{pdf}
\end{figure} 

Figure~\ref{pdf} shows the distributions functions for the
translational velocities along the horizontal direction.  For sake of
comparison, all the velocities were rescaled by their mean square
values.  We notice that the two transversal velocity distribution
functions deviate from a Gaussian and are fitted by $f(c)=A/(exp(\beta
c^{\alpha}) +exp(-\beta c^{\alpha}))$, with: $ A=1.068$,
$\beta=1.74$, $\alpha=0.918$ .

The exponential tails appear to have a smaller slope than the
theoretically predicted 3/2 value, for a uniform system stochastically
driven~\cite{vannoije}. We notice the phenomenon already reported
in~\cite{Pagnani}: the rescaled velocity distribution of lighter
species display higher tails.

Figure~\ref{pdf} also shows the angular velocity distributions
(rescaled by their mean square value, see caption).

Our simulations show that such distributions have pronounced
non-Gaussian tails. The angular velocity distributions can be
described by the same scaling function we used for translational
velocity distributions.

\subsection{Collision time distribution}

\begin{figure}
\includegraphics[clip=true,width=10.0cm, keepaspectratio,angle=0]{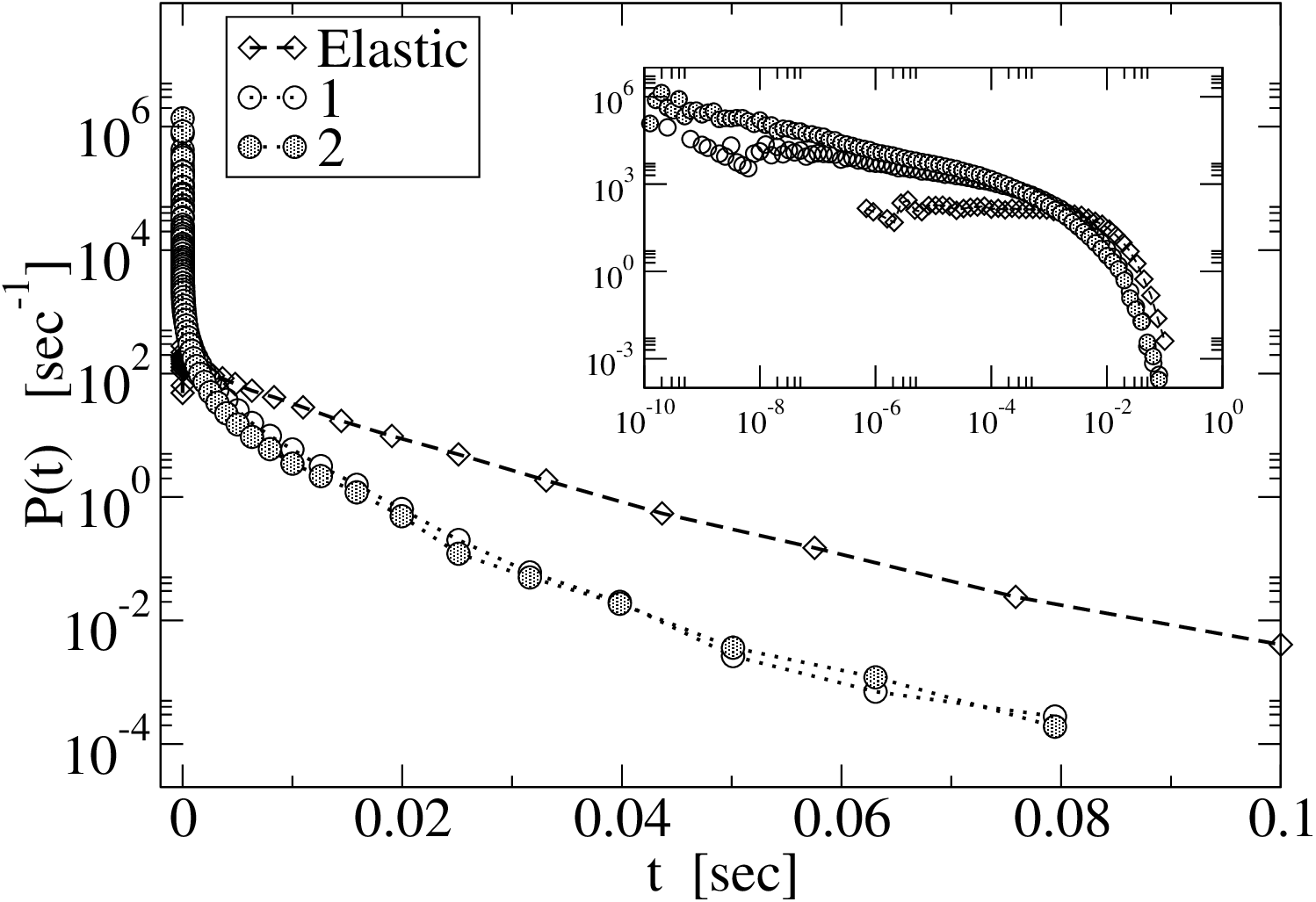}
\caption
{Probability distribution functions of the collision times for the two
components (system with $r_{11}=0.93$, $r_{22}=0.7$, $r_{12}=0.8$). For
comparison we plotted the corresponding distributions for an elastic
system. The inset shows the same curves in bilogarithmic scale.}
\label{tempi}
\end{figure}

Recently Blair and Kudrolli utilizing high speed digital photography
measured the collision statistics of grains bound to move on an
inclined plane~\cite{blair_poster}.  They determined the distribution of path
lengths, $P(l)$ and showed that it deviates from the theoretical
prediction for elastic hard spheres. In particular, $P(l)$ shows a
peak in the small $l$ region, not present in elastic systems. In order
to assess the existence of such a behavior in the IHS model we
performed similar measurements.  Figure~\ref{tempi} displays the
results for the collision times of the two species. One clearly sees
that the probability density that a particle suffers a collision in a
short interval is enhanced with respect to the elastic case. The
physical reason for that lies in the existence of strong correlations
which lead to the presence of clusters where the path lengths are
shorter than in a uniform system. For the sake of comparison we
plotted in figure~\ref{tempi} also the corresponding distribution of
an elastic gas having the same density and granular temperature of our
inelastic mixture. The effect of the shorter average collision time is
clearly visible. We also notice that the ratio of the collision
frequencies is approximately equal to the the ratio of the average
velocities, as expected from elementary kinetic arguments.

\begin{figure}
\includegraphics[clip=true,width=10.0cm, keepaspectratio,angle=-90]
{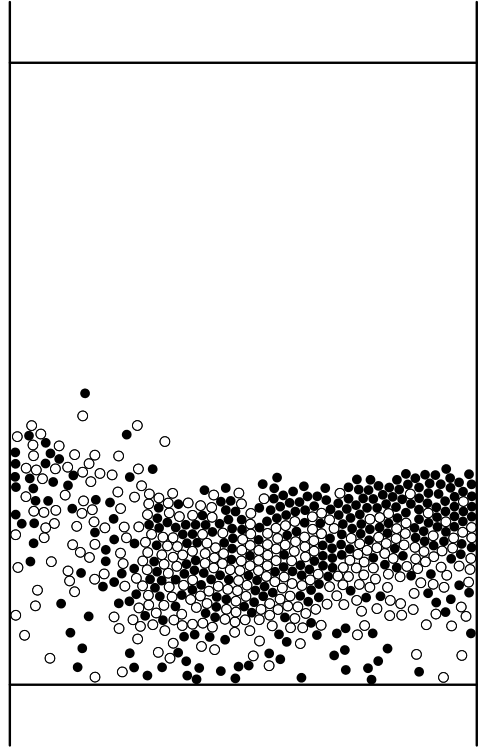}
\caption
{Snapshot of the system described in IV.
Open circles indicate particles of species 1, black circles particles
of species 2.}
\label{snapshot2}
\end{figure}

\section{Long system}

\begin{figure}
\includegraphics[clip=true,width=10.0cm, keepaspectratio,angle=0]{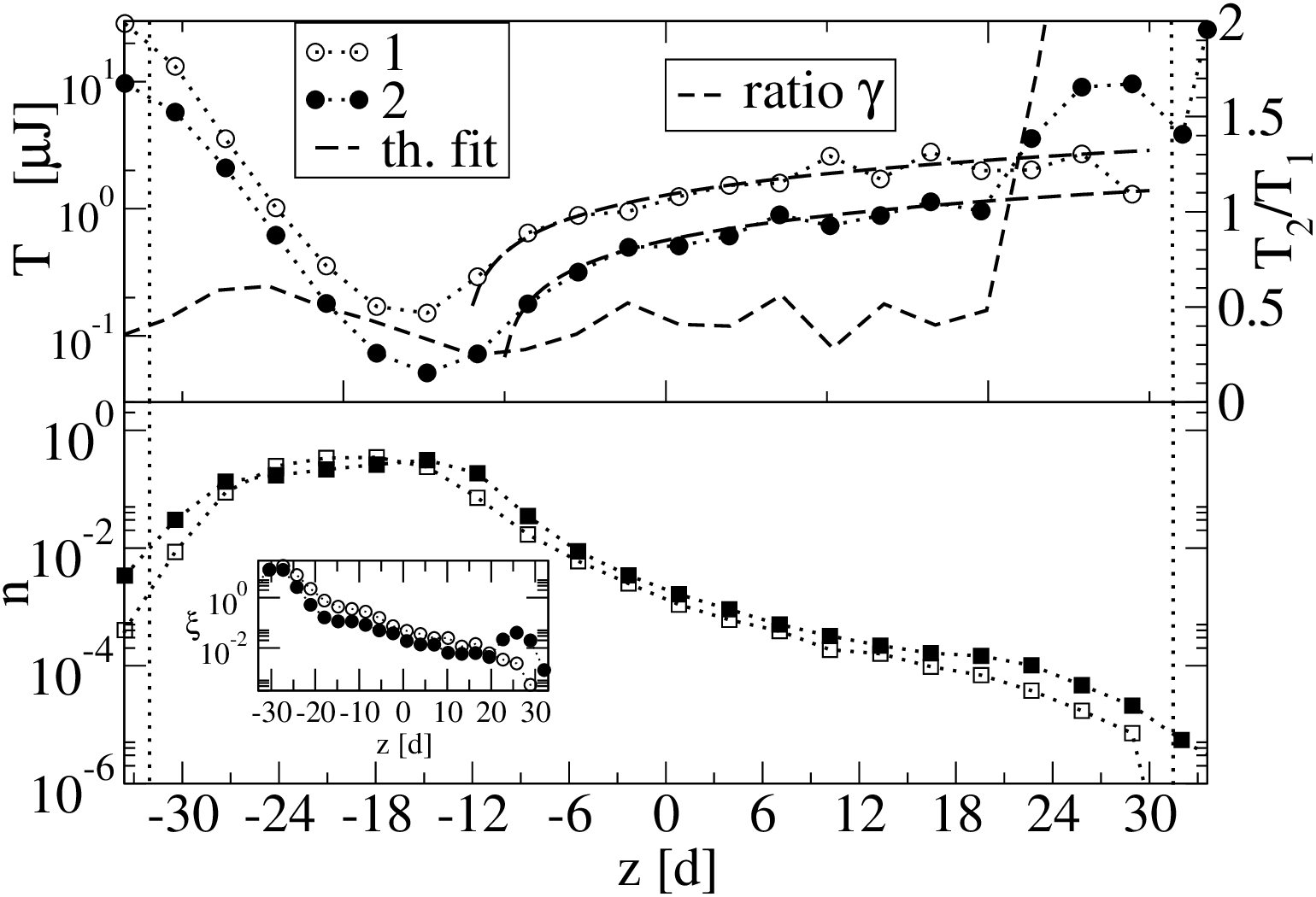}
\caption
{Temperature and area fraction profiles for the system discussed in
section IV. In the inset the dissipation rate $\xi(z)$ is shown. In
the figure we compared our result against the hydrodynamic prediction
$T \sim z^{3/2}$ (solid lines). The vertical dotted lines correspond
to the rest position of the horizontal boundaries.}
\label{lungo}
\end{figure}

In the case of a box of dimensions $L_x=48\,d$ and $L_y=64\,d$ with
$300$ grains of each species the effect of gravity is much more
evident, resulting in a stronger inhomogeneity and asymmetry of the
system. In fact, the density and temperatures profiles are not
symmetric with respect to the vertical direction.
A configuration of such system is shown in 
fig.~\ref{snapshot2}. Most of the
particles remain suspended above the bottom wall and are hit by those
which are between the bottom wall and and the bulk. Very few particles
reach the upper vibrating wall, so that the granular temperature of
the system is much lower in the top than in the bottom. The partial
temperature profiles and density profiles are shown in
fig. \ref{lungo}.  One sees that near the lower vibrating wall the
temperature profiles are similar to those of the shorter system,
whereas in the bulk they are appreciably different. The average ratio
is lower than in the $L_y=32$ d case.  An interesting feature present
in fig.\ref{lungo} is the presence of a region of increasing
temperatures. Such a phenomenon has been predicted by the hydrodynamic
theory of Brey et al.~\cite{Brey}.  The theory predicts a temperature
varying as $T \sim z^{3/2}$ in the region above the minimum. Such a
prediction is verified by our simulation results.

Physically, the increase is due to the competition between the energy
input from the walls and the small dissipation due to the reduced
number of collisions associated with the low density region. This can
be appreciated in the inset of figure~\ref{lungo} where the quantity
$\xi(z) = n(z)T(z)^{3/2}$ is shown: $\xi(z)$ is proportional to the
energy dissipation due to inelastic collisions among particles, being
the collision rate $\propto nT^{1/2}$ and the average dissipated
energy in a single collision $\propto T$. The graph of $\xi(z)$
reveals that the energy dissipation is much more relevant in the
bottom region than in the middle and upper regions.

\begin{figure}
\includegraphics[clip=true,height=10.0cm, keepaspectratio,angle=-90]{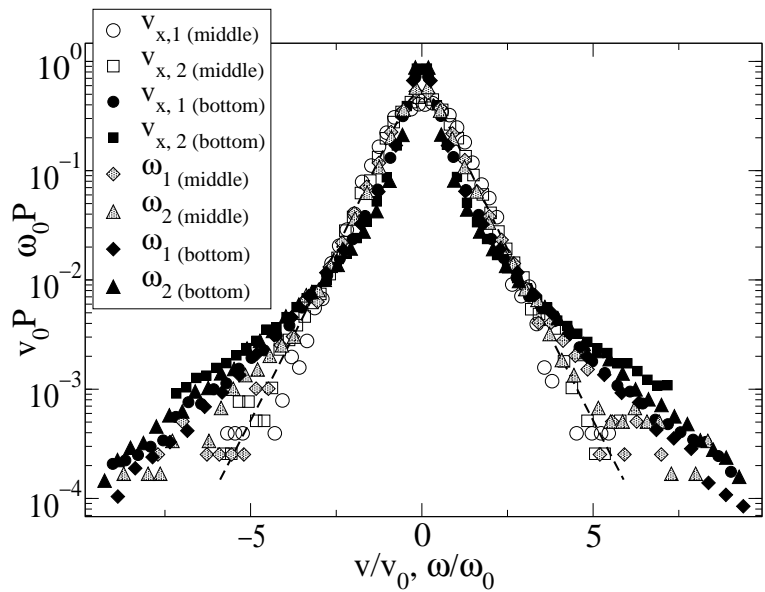}
\caption
{Rescaled velocity distribution functions for the $48 \times 64$ system
measured at different heights.}
\label{pdf_lungo}
\end{figure}

Correspondingly the velocity PDF shows an interesting behavior (see
figure~\ref{pdf_lungo}).  In fact, the shape of the rescaled PDF
becomes narrower with the height. In the central region the measured
PDF resembles the exponential PDF measured in the short system
(figure~\ref{pdf}), whereas in the bottom region the PDF has more
extended tails. Such a lack of universality in the velocity PDf was
noted experimentally by Blair and Kudrolli \cite{blair}. It was also
noticed in~\cite{puglisi1}: there it was shown that the tails of the
PDF became broader when the dissipation rate was increased.  Here the
mechanism is similar: the broader PDFs are those measured in the
bottom region, where the dissipation rate is higher.

\section{Conclusion}

Summarizing, we studied a systems of inelastic particles in vertically
vibrated containers and subject to the gravitational field. The system
was numerically investigated by using an Event-Driven dynamics, which
affords the exploration of a wide range of dynamical parameters. We
examined two similar setups which differ only for the aspect ratio
of the container and for the number of particles. 

In the shorter system we determined the partial granular
temperatures, their ratio, and the area fraction profiles along the
vertical direction. Those measures are in qualitative agreement with
the experimental results, but a quantitative comparison requires a
more detailed knowledge of the experimental parameters.  Physically
the lack of energy equipartition between the two species has two
causes: a) the vibrating walls feed energy proportionally to the mass
of each species; b) the dissipation is in general different for the
two components. The two effects may conspire (for example if the
heavier species is the more elastic) to give a small value of the
temperature ratio or they may cancel and give a temperature ratio
close to one (for example when the heavier species is the more
inelastic).  Moreover, we observed that density profiles are
non-uniform along the horizontal direction, as well, indicating that
the particles tend to clusterize in the vicinity of side walls.  We
also measured the velocity distributions, verifying that they can be
collapsed onto each other, under proper rescaling, and the scaling
function has a stretched exponential behavior. Finally the
distribution of flight times between succesive collisions has been
measured and compared to that of an elastic system: the inelasticity
has the effect of enhancing the statistics of very short times.

In the longer system we have again obtained the temperature and area
fraction profiles, observing a stronger inhomogeneity and
asymmetry. In this case the area fraction is much larger
near the bottom wall, reaching higher values than in the previous
experiment. The middle-upper region seems to be qualitatively well
described by recent hydrodynamic theories developed for one component
systems. Here the velocity PDFs display a non-universal behavior, with
broader tails in the more dense (and dissipative) regions.



\section{Appendix}

In order to make it simpler for the reader to interpret the present
model, we present an appendix in which we explicitly state the
collision rules\cite{Farkas}.

The colliding particles are characterized by radii $R_1$ and $R_2$,
positions ${\bf r}_1$ and ${\bf r}_2$, translational velocities ${\bf
v}_1$ and ${\bf v}_2$ and rotational (angular) velocities
$\boldsymbol{\omega}_1$ and $\boldsymbol{\omega}_2$ (we assume that if
$\boldsymbol{\omega}$ is parallel and in the direction of the $z$
axis, than the rotation is anticlockwise if seen from above the $xy$
plane). We introduce the normal unitary vector joining the centers of
the particles ${\bf n} = ({\bf r}_2 - {\bf r}_1)/|{\bf r}_2 - {\bf
r}_1|$ and the tangential unitary vector ${\bf t}$ obtained rotating
${\bf n}$ by an anti-clockwise angle $\pi/2$. Then we introduce the
relative velocity ${\bf g}$, the velocity of the center of mass ${\bf
V}$ and the velocities of the particles in the center of mass frame
$\boldsymbol{\zeta}_1$ and $\boldsymbol{\zeta}_2$:

\begin{subequations}
\begin{align}
{\bf g}&={\bf v}_1 - {\bf v}_2\\
{\bf V}&=\frac{m_1{\bf v}_1 + m_2{\bf V}_2}{m_1+m2}\\
\boldsymbol{\zeta}_1&=\frac{m_{eff}}{m_1}{\bf g}\\
\boldsymbol{\zeta}_2&=-\frac{m_{eff}}{m_2}{\bf g}
\end{align}
\end{subequations}
where $m_{eff}=m_1m_2/(m_1+m_2)$.

Then we decompose the relative velocity ${\bf g}$ on the orthonormal
basis given ${\bf n}$ and ${\bf t}$, as well as the velocities of the
particles in the center of mass frame, i.e.:

\begin{subequations}
\begin{align}
g_n & = ({\bf v_1} - {\bf v_2})\cdot{\bf n} \\
g_t & = ({\bf v_1} - {\bf v_2})\cdot{\bf t} \\
\zeta_{n\beta} &= \frac{m_{eff}}{m_\beta} g_n \\
\zeta_{t\beta} &= \frac{m_{eff}}{m_\beta} g_t 
\end{align}
\end{subequations}
with $\beta=1,2$ the index of the particle.

We finally introduce $g_c$ as the relative circular velocity at the
point of contact and $g_r$ as the total tangential relative velocity
(circular and translational) at the point of contact:

\begin{subequations}
\begin{align}
g_c & = R_1\omega_1 + R_2\omega_2\\
g_r & = g_c + g_t
\end{align}
\end{subequations}

To characterize the collision rules we use a model that take into account a
reduction of normal relative velocity ($g_n$), a reduction of total
tangential relative velocity ($g_r$) and an exchange of energy between
those two degrees of freedom. The reduction of normal relative
velocity is modeled as usual by means of a restitution coefficient
$\alpha \in [0,1]$:

\begin{equation}  \label{normal}
g_n^\prime = -\alpha g_n.
\end{equation}
We assume a dependence of $\alpha$ by the relative velocity of the form:

\begin{displaymath}
\alpha (g_n) = \left\{ 
\begin{array}{lll}
1 - (1 - \alpha_0) \frac{|g_n|}{v_0}^\frac{3}{4} & \mbox{ per } & g_n < v_0 \\
                                        \alpha_0 & \mbox{ per } & g_n > v_0 
\end{array}
\right.
\end{displaymath}
where $\alpha_0$ is a constant which stands for $r_{ij}$, $v_0
=\sqrt{gD}$, with $D$ the average diameter of the particles (and $g$
is the gravity acceleration). From equations~\eqref{normal} we obtain
the update of normal velocities in the center of mass frame:

\begin{subequations}
\begin{align}
\zeta_{n1}^\prime & =  -\alpha \frac{m_{eff}}{m_1} g_n\\
\zeta_{n2}^\prime & = \alpha \frac{m_{eff}}{m_2} g_n\nonumber
\end{align}
\end{subequations}

What lacks now is an expression for the tangential and angular
velocities after the collisions. We distinguish between two possible
cases: sliding or sticking collisions. The condition that allows to
determine if a collision is sticking or sliding is the following:

\begin{equation}
\frac{|g_r|}{g_n} \geq \frac{l+1}{l}\mu(1 + \alpha)
\end{equation}
where $l$ is the adimensionalized inertia moment and is equal to $1/2$
or $2/5$ if the particle is a disk or a sphere respectively, while
$\mu$ is the static friction coefficient of the surface of the
particles which in the following will be assumed to be equal to the
dynamic friction coefficient.

In the sliding case we use the following rules to update the
tangential components of the velocities of the particles in the center
of mass frame:

\begin{subequations}
\begin{align}
\zeta_{t1}^\prime & =  \zeta_{t1} - \mu(1 + \alpha)\frac{m_{eff}}{m_1}g_n sign(g_r)\\
\zeta_{t2}^\prime & =  \zeta_{t2} + \mu(1 + \alpha)\frac{m_{eff}}{m_2}g_n sign(g_r)\\
R_1\omega_1^\prime & = R_1\omega_1 - \frac{\mu(1 + \alpha)}{l}\frac{m_{eff}}{m_1}g_n sign(g_r)\\
R_2\omega_2^\prime & = R_2\omega_2 - \frac{\mu(1 + \alpha)}{l}\frac{m_{eff}}{m_2}g_n sign(g_r)
\end{align}
\end{subequations}

In the case of a sticking collision, instead, the update rules are
obtained considering that:

\begin{equation}
\zeta_{t1}^\prime - \zeta_{t2}^\prime + R_1\omega_1^\prime + R_2\omega_2^\prime = 0
\end{equation}
from which, after calculations, one gets:

\begin{subequations}
\begin{align}
\zeta_{t1}^\prime & =  \frac{1}{l + 1}\zeta_{t1} - \frac{l}{l+1}\frac{m_{eff}}{m_1}(R_1\omega_1 + R_2\omega_2)\\
\zeta_{t2}^\prime & =  \frac{1}{l + 1}\zeta_{t2} + \frac{l}{l+1}\frac{m_{eff}}{m_2}(R_1\omega_1 + R_2\omega_2)\\
R_1\omega_1^\prime & =  R_1\omega1\left[\frac{l}{l+1}+\frac{m_{eff}}{(l+1)m_2}\right]\\
& -  R_2\omega_2\frac{m_{eff}}{(l+1)m_1} - \frac{1}{l+1}\zeta_{t1}\\
R_2\omega_2^\prime & =  R_2\omega2\left[\frac{l}{l+1}+\frac{m_{eff}}{(l+1)m_1}\right]\\
& -  R_1\omega_1\frac{m_{eff}}{(l+1)m_2} + \frac{1}{l+1}\zeta_{t2}
\end{align}
\end{subequations}

The velocity of particles in the absolute frame are finally obtained
in the two cases (considering that the center of mass is not perturbed by the collision) by the equation:

\begin{equation}
\mathbf{v}^\prime_{\beta}=\mathbf{V}+\zeta_{n\beta}^\prime \mathbf{n} + \zeta_{t\beta}^\prime \mathbf{t}
\end{equation}
(for particle of index $\beta$) leading to the following global
collision rule for translational velocities:

\begin{subequations}
\begin{align}
\mathbf{v}_1^\prime&=\mathbf{v}_1-(1+\alpha)\frac{m_2}{m_1+m_2}[(\mathbf{v}_1-\mathbf{v}_2)\cdot \mathbf{n}]\mathbf{n}-
\left\{ 
\begin{array}{lll}
sign(g_r)\mu(1+\alpha)\frac{m_2}{m_1+m_2}[(\mathbf{v}_1-\mathbf{v}_2)\cdot \mathbf{n}]\mathbf{t} \mbox{  (sliding)} \\
\frac{m_1-lm_2}{(l+1)(m_1+m_2)}[(\mathbf{v}_1-\mathbf{v}_2)\cdot \mathbf{t}]\mathbf{t} - \frac{l}{l+1}\frac{m_2}{m_1+m_2}(R_1\omega_1 + R_2\omega_2) \mbox{  (stick)} 
\end{array}
\right. \\
\mathbf{v}_2^\prime&=\mathbf{v}_2+(1+\alpha)\frac{m_1}{m_1+m_2}[(\mathbf{v}_1-\mathbf{v}_2)\cdot \mathbf{n}]\mathbf{n}+
\left\{ 
\begin{array}{lll}
sign(g_r)\mu(1+\alpha)\frac{m_1}{m_1+m_2}[(\mathbf{v}_1-\mathbf{v}_2)\cdot \mathbf{n}]\mathbf{t} \mbox{  (sliding)} \\
\frac{m_2-lm_1}{(l+1)(m_1+m_2)}[(\mathbf{v}_1-\mathbf{v}_2)\cdot \mathbf{t}]\mathbf{t} + \frac{l}{l+1}\frac{m_1}{m_1+m_2}(R_1\omega_1 + R_2\omega_2) \mbox{  (stick)} 
\end{array}
\right.
\end{align}
\end{subequations}

while for rotational velocities:

\begin{subequations}
\begin{align}
R_1\omega_1^\prime & = 
\left\{ 
\begin{array}{lll}
R_1\omega_1 - \frac{\mu(1 + \alpha)}{l}\frac{m_2}{m_1+m_2}g_n sign(g_r) \mbox{  (stick)} \\
R_1\omega1\left[\frac{l}{l+1}+\frac{m_1}{(l+1)(m_1+m_2)}\right] -  R_2\omega_2\frac{m_2}{(l+1)(m_1+m_2)} - \frac{1}{l+1}\zeta_{t1} \mbox{  (slide)}\\
\end{array}
\right.\\
R_2\omega_2^\prime & = 
\left\{ 
\begin{array}{lll}
R_2\omega_2 - \frac{\mu(1 + \alpha)}{l}\frac{m_1}{m_1+m_2}g_n sign(g_r) \mbox{  (stick)} \\
R_2\omega_2\left[\frac{l}{l+1}+\frac{m_2}{(l+1)(m_1+m_2)}\right] -  R_1\omega_1\frac{m_1}{(l+1)(m_1+m_2)} + \frac{1}{l+1}\zeta_{t1} \mbox{  (slide)}\\
\end{array}
\right.
\end{align}
\end{subequations}

\begin{acknowledgments}
This work was supported by Ministero dell'Istruzione,
dell'Universit\`a e della Ricerca, Cofin 2001 Prot. 2001023848,
by INFM and INFM {\it Center for Statistical Mechanics and Complexity}
(SMC).
\end{acknowledgments}


\begin{thebibliography}{0}

\bibitem{general} 
H.M. Jaeger, S.R. Nagel and
R.P. Behringer, {\em Rev. Mod. Phys.} {\bf 68}, 1259 (1996).

\bibitem{Zanetti}I.Goldhirsch and G.Zanetti, 
Phys.Rev.Lett. {\bf 70}, 1619 (1993).

\bibitem{Brey} J.J.Brey, M.J. Ruiz-Montero, F. Moreno,
R. Garcia-Rojo, cond-mat/0201435 

\bibitem{Baldassa}A. Baldassarri, U. Marini Bettolo Marconi 
and A. Puglisi, Phys.Rev. E {\bf 65}, 051301 (2002).

\bibitem{Feitosa} 
K. Feitosa e N. Menon, 2002,Phys.Rev. Lett. {\bf 88}, 198301 (2002). 

\bibitem{Parker}
R.D. Wildman and D.J. Parker, Phys.Rev. Lett. {\bf 88}, 064301 (2002).

\bibitem{dufty2}
S.R. Dahl, C.M. Hrenya,
V. Garz\'o e J. W. Dufty,2002, cond-mat/0205413.

\bibitem{dufty}
Vicente Garz\'o e James Dufty, Phys. Rev. E, {\bf 60} 5706 (1999).

\bibitem{ioepuglio}
U. Marini Bettolo Marconi and A. Puglisi,
Phys.Rev. E {\bf 65}, 051305 (2002).


\bibitem{ioepuglio2}U. Marini Bettolo Marconi and
A. Puglisi, Phys. Rev. E {\bf 66}, 011301 (2002).

\bibitem{Barrat} 
Alain Barrat e Emmanuel Trizac, 2002, cond-mat/0202297.

\bibitem{Pagnani}
A.Pagnani, U. Marini Bettolo Marconi, and A. Puglisi,
(cond-mat/0205619).

\bibitem{barrat2} 
A. Barrat e E. Trizac, 2002, cond-mat/0207267.
 
\bibitem{Walton}
O.R. Walton, Particulate Two-Phase Flow, M.C. Roco editor,
Butterworth-Heinemann, Boston (1993), page 884.

\bibitem{Bizon} 
C. Bizon, M. D. Shattuck, J. B. Swift and H. Swinney, Phys. Rev. E
{\bf 60}, 4340 (1999)

\bibitem{Cattuto} 
C. Cattuto, submitted to Computers in Science and Engineering.

\bibitem{puglisi1}
A. Puglisi, V. Loreto, U. Marini Bettolo Marconi, A. Petri and
A. Vulpiani, 1998 Phys. Rev. Lett. {\bf 81}, 3848 and Phys. Rev. E
{\bf 59}, 5582 (1999).

\bibitem{Luding}
S. McNamara and S.Luding, Phys.Rev. E {\bf 58}, 2247 (1998).

\bibitem{Baldassa_old}
A. Baldassarri, U. Marini Bettolo Marconi, A. Puglisi, A. Vulpiani
{\em Phys. Rev. E} {\bf 64}, 011301 (2001).

\bibitem{vannoije} 
T. P. C. van Noije and M. H. Ernst, Granular Matter {\bf 1}, 57 (1998).

\bibitem{blair_poster}
D.L. Blair and A. Kudrolli, private comunication.

\bibitem{blair}
D.L. Blair and A. Kudrolli, Phys. Rev. E {\bf 64}, 050301 (2001).

\bibitem{Farkas} Z. Farkas, P. Tegzes, A. Vukics, T. Vicsek,
cond-mat/9905094 version 7/5/1999.

\end{thebibliography}
\end{document}